\documentstyle{article}
\begin{document}
\begin{center}
{\Large\bf An appropriate candidate for exact distribution of
closed random walks using quantum groups}
\end{center}
\begin{center}
\textbf{S. A. Alavi $^{\dag}$, M. Sarbishaei}\\

\textit{Department of Physics, Ferdowsi University of Mashhad,
Mashhad, P. O. Box 1436, Iran}
\end{center}

\textbf{Abstract.}\emph{We show that the structure of the quantum
group $su_{q}(2)$ is intimately related to the random walks on a
two dimentional lattice. Using this connection we obtain an
appropriate candidate for the exact area distribution of closed
random walks of length $N$ on a two dimensional square lattice. We
compare our results with exact enumeration.}\\



\textbf{1.Introduction.} Let us consider a spinless electron on a
two dimensional lattice and submitted to a uniform magnetic field
along the z-direction and perpendicular to the plane of motion.
The Hamiltonian of the system is:
\begin{equation}
h=\frac{1}{2m}(\vec{p}+e\vec{A})^{2}.
\end{equation}
The system is not invariant under translations but there is an
invariance under the so-called magnetic translation operators [1]:
\begin{equation}
w(\vec{a})=\exp(\frac{i}{\hbar}\vec{a}.\vec{k}) ,
\end{equation}
where $\vec{a}$ is an arbitrary two dimensional lattice vector and
$\vec{k}$ is:
\begin{equation}
\vec{k}=k_{x}\hat{x}+k_{y}\hat{y}=\vec{p}+e\vec{A}+e\vec{r}\times\vec{B}.
\end{equation}
There is a relationship between magnetic translation operators,
quantum groups, Landau levels and the quantum Hall effect which
has been the subject of  study in several references [e.g.,2-4]. It
is shown that $w(a)$ satisfy the following relation:
\begin{equation}
w(\vec{a}) w(\vec{b})=\exp(i\frac{e}{2\hbar}\vec{B}.(\vec{a}\times\vec{b}))w(\vec{a}+\vec{b}) ,
\end{equation}
where $\vec{a}\times\vec{b}=a_{1}b_{2}-a_{2}b_{1}$,which is exactly the
q-commutation relation:
\begin{equation}
w(\vec{a}) w(\vec{b})= q w(\vec{b}) w(\vec{a}) ,
\end{equation}
$\overline{\dag.Corresponding}$ $\overline{author.}$\\
E-mail:$Alavi@sttu.ac.ir$ and  
       $s_{-} alialavi@hotmail.com$\\

where $q=\exp(i\frac{e}{\hbar}\vec{B}.(\vec{a}\times\vec{b}))$. We can write $q$ in the form $q=e^{i\gamma}$ where $\gamma=2\pi\frac{\Phi}{\Phi_{0}}$.
$\Phi$ and $\Phi_{0}$ are the magnetic flux through the unit cell and the quantum of the flux respectively :\\

\begin{equation}
\Phi=\vec{B}.(\vec{a}\times\vec{b}),  \Phi_{0}=\frac{h}{e} .
\end{equation}
Here $\vec{a}$ and $\vec{b}$ are two perpendicular unit vectors which build the lattice  unit cell, $\mid a\mid=\mid b\mid=1$.
 Using (4) and (5) it is easy to show that the following
combinations of the magnetic translation operators :
\begin{equation}
j_{+}=\frac{w(\vec{a})+w(\vec{b})}{q-q^{-1}} ,
\end{equation}
\begin{equation}
j_{-}=-\frac{w(\vec{-a})+w(\vec{-b})}{q-q^{-1}} ,
\end{equation}
and $q^{j_{3}}=w(\vec{b}-\vec{a})=j_{z}$ form the $su_{q}(2)$ algebra. 
The $su_{q}(2)$ algebra is the q-deformation of the lie algebra
$su(2)$ [5-8], with generators satisfying the commutation relations:
\begin{equation}
[j_{3},j_{\pm}]=\pm j_{\pm}.
\end{equation}
\begin{equation}
[j_{+},j_{-}]=\frac{q^{2j_{3}}-q^{-2j_{3}}}{q-q^{-1}} .
\end{equation}
We observe that (4), (5) and the $su_{q}(2)$ algebra are invariant
under following transformations:
\begin{equation}
w(\vec{a})\rightarrow g(q) w(\vec{a}) ,
\end{equation}
\begin{equation}
w(-\vec{a})\rightarrow g^{-1}(q) w(-\vec{a}) ,
\end{equation}
or equivalently:
\begin{equation}
j_{+}\rightarrow g(q) j_{+},
\end{equation}
\begin{equation}
j_{-}\rightarrow g^{-1}(q) j_{-},
\end{equation}
\begin{equation}
j_{z}\rightarrow j_{z},
\end{equation}
where $g(q)$ is an arbitrary function of $q$.\\

\textbf{2. Areas distribution of closed random walks.}

Bellissard et al [9] derived the distribution of areas of closed
random walks using noncommutative geometry. Their method was based on the Harper model. The Harper model Hamiltonian [10] is given by:
\begin{equation}
H=\sum_{\mid a\mid =1}w(\vec{a})=w(\vec{a})+w(\vec{b})+w(-\vec{a})+w(-\vec{b}) .
\end{equation}
We define the trace per unit area such that
\begin{equation}
T({w(\vec m_{1}}){w(\vec m_{2}}))=\delta_{\vec{m}_{1}+\vec{m}_{2},0} .
\end{equation}
Using (16) and (17) it is easy to see that :
\begin{equation}
T(H^{N})=\sum_{\Gamma}e^{\frac{i\gamma}{2}A(\Gamma)} ,
\end{equation}
where the sum is over the set of closed paths starting at the origin of length $N$, and $A(\Gamma)$ is the algebraic area closed by $\Gamma$.
If $\Omega_{N}$ is the number of such closed paths, we have [9]:
\begin{equation}
\sum_{A=-A_{max}}^{A_{max}}P_{N}(\frac{A}{N})\exp(\frac{ixA}{N})=\Omega_{N}^{-1}\sum_{\Gamma}
\exp(\frac{ixA(\Gamma)}{N})=\Omega_{N}^{-1}T(H^{N})\mid_{\gamma=\frac{x}{N}}
,
\end{equation}
where $A_{max}$ is the maximum algebraic area closed by $\Gamma$ with nonvanishing probability. Our aim is to find the formula for $P_{N}(a)$ where $a$ is renormalized area $a=\frac{A}{N}$, and to investigate its behaviour. \\ 
From (7), (8) and (16) we obtain:
\begin{equation}
H=w(\vec{a})+w(\vec{b})+w(\vec{-a})+w(\vec{-b})=2i(q-q^{-1})j_{y} .
\end{equation}
Thus we observe that the Harper Hamiltonian operator is a
generator of the $su_{q}(2)$ algebra. If the eigenvalue of $j_{3}$
is
$m$, the eigenvalue of $j_{z}$ will be $\pm q^{m}$. Using (17) or its equivalent :
\begin{equation}
T(w(c))=\delta_{c,0} .
\end{equation}
one can show that:
\begin{equation}
T(w(\vec{b}-\vec{a}))=T(\sum_{a}w(\vec{a}))=T(j_{y})=0 ,
\end{equation}
This result also follows from (9) and (10) directly. A
straightforward calculation show that the eigenvalues of $H$ have
exponential form(see for example equ.(18)). To obtain the
eigenvalues of $H$,we note that:\\
1. $E$ depends on the quantum number $m$, and is a function of q-deformation parameter $\gamma$(actualy $\gamma$ and $N$ are not independent, $\gamma=\frac{x}{N}$, see equ.(19), and $E$ can be a function of both of them). It is also has exponential form ( for more details see appendix A) and
satisfies in equ.(22).\\
From mathematical point of view, if a series of functions
\begin{equation}
\sum_{n=1}^{\infty}f_{n}(x).
\end{equation}
is convergent for any $x$ in some interval, then this series defines a function of variable $x$,
$f(x)=\sum_{n=1}^{\infty}f_{n}(x)$ in this interval. The importance of a series (e.g.
a solution of some differential equations) lies in its
convergence. For example, if $\ell$ be an integer then the series
expansions of the solution of the Legendre equation
$P_{\ell}(x)$ and $Q_{\ell}(x)$ are convergent, and they convert to 
polynomials. The same statement is also true for the hypergeometric functions and the other polynomials.\\ We know that the energy of a physical system must be finite, therefore the series (6) in [9] is convergent and is a function of $\gamma$ (or $N$) which we will try to determine it.\\
On the other hand we know that the operators $j_{y}$ and $j_{z}$
of $su(2)$ algebra do not commute and can not have common
eigenvectors, but their eigenvalues are the same. After deformation the eigenvalues of the $j_{y}$ and $j_{z}$ will be not the same, but still their eigenvalues will be discrete labelled by integer quantum number $m$.\\
2. We know the eigenvalues of $H$, in the limit of large $N$ [9].\\
Therefore we can write :
\begin{equation}
E_{m}=\pm e^{u_{m}(\gamma)} .
\end{equation}
If we choose $u_{m}(\gamma)=-m\frac{\gamma}{4}+h(\gamma)$, then we
have:
\begin{equation}
E_{m}=\pm f(\gamma) e^{-m\frac{\gamma}{4}} ,
\end{equation}
where $f(\gamma)=\exp (h(\gamma))$. As the candidate for $f(\gamma)$ we take :
\begin{equation}
f(\gamma)=2(e^{i(\frac{\frac{\gamma}{4}}{\sinh(\frac{N\gamma}{4})})}+e^{-i(\frac{\frac{\gamma}{4}}{\sinh(\frac{N\gamma}{4})})})
.
\end{equation}
We show that its $N\rightarrow\infty$ is fully consistent with the result of [9]. We tested it numerically(see below).  
Then we will have:
\begin{equation}
E_{m}=\pm 4\cos(\frac{\frac{\gamma}{4}}{\sinh(\frac{N\gamma}{4})})
e^{-\frac{m\gamma}{4}} ,
\end{equation}
By introducing the new variable $\ell$
\begin{equation}
m=2\ell +1 ,
\end{equation}
We have:
\begin{equation}
E_{\ell}=\pm
4\cos(\frac{\frac{\gamma}{4}}{\sinh(\frac{N\gamma}{4})})e^{-\frac{(2\ell
+1)\gamma }{4}} .
\end{equation}
Expanding $E_{\ell}$ for small values of $\gamma$,
$\gamma\rightarrow 0$, we obtain:
\begin{equation}
E_{\ell}=\pm 4 [(1-\frac{1}{2N^{2}})-(2\ell +1)\frac{\gamma}{4}+[(2\ell
+1)^{2}+1]\frac{\gamma^{2}}{32}-O(\gamma^{3})] ,
\end{equation}
The term $\frac{1}{2N^{2}}$ can be omitted, because it tends to zero when $N\rightarrow\infty$ ( $\gamma\rightarrow 0$), and equ.(30) reproduces the eigenvalues used by Bellissard[9]. Now $T(H^{N})$ is
given by:
\begin{equation}
T(H^{N})=\sum_{\pm}\sum_{\ell}
(E_{\ell}^{\pm}(\gamma))^{N}(\frac{\gamma}{2\pi}) ,
\end{equation}
where $\gamma=\frac{x}{N}$ is the multiplicity per unit area. This
leads to:
\begin{equation}
T(H^{N})=\frac{4^{N+1}}{2\pi
N}\frac{\frac{x}{4}}{\sinh(\frac{x}{4})}
\cos^{N}(\frac{\frac{x}{4}}{N\sinh(\frac{x}{4})}) ,
\end{equation}
for even $N$. This gives us the exact characteristic function for
the distribution of area. In the limit of large $N$ we obtain:
\begin{equation}
T(H^{N}(x))=\frac{4^{N+1}}{2\pi
N}\frac{\frac{x}{4}}{\sinh(\frac{x}{4})}[1-\frac{1}{2N}\frac{(\frac{x}{4})^{2}}{\sinh^{2}(\frac{x}{4})}+O(\frac{1}{N^{2}})]
.
\end{equation}
which is also derived by Bellissard [9]. From (19) and (32) we
have:
\begin{equation}
\Omega_{N}=T(H^{N})\mid_{\gamma=0}=\frac{4^{N+1}}{2\pi
N}\cos^{N}(\frac{1}{N}) ,
\end{equation}
This gives the total number of closed paths of length N. In the
limit of large $N$  we have:
\begin{equation}
\Omega_{N}=\frac{4^{N+1}}{2\pi N}(1+(O\frac{1}{N})) .
\end{equation}
Using (19), (32), (34) and the normalization condition we can
obtain a series expansion for the probability distribution: \\

$P_{N}(a)=\frac{4}{\pi
N}\sum_{m=0}^{\infty}\sum_{k=0}^{\infty}\sum_{\ell=0}^{m+1}$
\begin{equation}
(-1)^{\ell} G_{m}^{N} 2^{2m+1}\\
(_{ k}^{2m+k}) (_{ 2\ell}^{2m+2})
\frac{(2m+2k+1)^{2m+2-2\ell}(16a^{2})^{\ell}}{((2m+2k+1)^{2}+16a^{2})^{2m+2}}
\Gamma(2m+2) ,
\end{equation}
We have plotted $P_{N}(a)$ as given by equ.(36) against exact
enumeration[11] in fig1. In the summation of equ.(36) we have
allowed $m$ to range from $0$ to $2$ and k from $0$ to $7000$ for
fig.1.(a) and from $0$ to $60000$ for fig.1.(b) and (c). Comparing
fig.1.(a) and (b), we find that a better result will be obtain if
we increase the range of summation over $k$. The same statement is
also true for $m$. $m=0$ and $m=1$ give the equ.(9) in [9] and the
$\frac{1}{N}$ correction term made by them respectively. Some of
the $G_{m}^{N}$ coefficients   are presented in Appendix B.\\
In conclusion we observe that the expression presented for the spectrum of $H$  using quantum groups is in good agreement with the results for $N\rightarrow\infty$ and also with exact enumeration. If there is other candidate for $f(\gamma)$, which full fills all the requirements, then the difference between it and our expression will be negligibly small.\\ 

\textbf{Acknowledgment.}
One of the authors(S.A.A.) would like to thank Professor S.
Rouhani for valuable discussions. He is also very grateful to
Professor P. Kulish and Professor P. Pre\v{s}najder for their careful reading of the manuscript and
for their valuable comments and finally I wish to thank K. Khabbazi
and O. Naser Ghodsi for their assistance on computational
aspects of this work.\\

\textbf{Appendix A.}
The Harper Hamiltonian is:
\begin{equation}
H=e^{a_{1}}+e^{a_{2}}+e^{-a_{1}}+e^{-a_{2}}=\sum_{a_{i}}e^{a_{i}}
,
\end{equation}
where $[a_{i},a_{j}]=u_{ij}(\gamma)$,
$e^{a_{i}}e^{a_{j}}=e^{a_{i}+a_{j}}e^{\frac{1}{2}[a_{i},a_{j}]}$
and $T(e^{a_{i}})=\delta_{a_{i},0}$.
then:
\begin{equation}
H^{N}=\sum_{a_{i}}\sum_{a_{j}}....\sum_{a_{N}}
e^{a_{i}}e^{a_{j}}....e^{a_{N}} .
\end{equation}
therefore we have:
\begin{equation}
T(H^{N})=\sum_{m}e^{u_{m}(\gamma)} .
\end{equation}
$u_{m}(\gamma)$ must be determined using boundary conditions.\\

\textbf{Appendix B.}  
The first four $G_{m}^{N}$ coefficients are as follows:
\begin{equation}
G_{0}^{N}=1
\end{equation}
\begin{equation}
G_{1}^{N}=-\frac{1}{2N} .
\end{equation}
\begin{equation}
G_{2}^{N}=\frac{1}{24}\frac{1}{N^{3}}+\frac{1}{8}\frac{N-1}{N^{3}}
.
\end{equation}
\begin{equation}
G_{3}^{N}=-\frac{1}{720}\frac{1}{N^{5}}-\frac{1}{48}\frac{N-1}{N^{5}}-\frac{1}{48}\frac{(N-1)(N-2)}{N^{5}}
.
\end{equation}\\

\textbf{Figure captions.}\\
Fig.1. The exact and approximate distribution for areas of closed random paths. The exact 
distribution is plotted for $N=16$. The solid points were obtained through exact enumerations [11].\\

\textbf{References.}\\
1. M.Chaichian and A.Demichev, Introduction to quantum groups,
pp. 183-185, World Scientific, Singapore (1996).\\
2. Haru-Tada Sato, Modern Physics Letters A, Vol.9, No.20,
1819-1825, 1994.\\
3. Haru-Tada Sato, Modern Physics Letters A, Vol.9, No.5, 451-458,
1994.\\
4. Alimohammadi and A. Shafei Deh Abad, J. Phys. A: Math. Gen. 29,
559-563, 1996.\\
5. P. Kulish and Yu. Reshtikhin, J. Sov. Math. 23 (1983) 2435 (translation 
from: Zapiski Nauch. Seminarov LOMI 101 (1981) 101.)  \\
6. G. Drinfeld in Proc. ICM Berkeley, CA, ed A. M.
Gleason(Providence, RI:AMS) P 798 1986.\\
7.M. Jimbo Lett. Math. Phys 10 63, 1985; Lett. Math. Phys 11 247,
1986.\\
8. M. Chaichian and P. Kulish, Phys. Lett. B 234 (1990) 72.\\
9. Jean Bellissard et al J. Phys. A: Math. Gen. 30(1997)
L707-L709\\
10. Harper P G 1955 Proc. Phys. Soc. Lond. A 68 874.\\
   Harper P G 1955 Proc. Phys. Soc. Lond. A 68 879.\\
11. R. Afsari, N. Sadeghi and S. Rouhani, Exact enumeration of
closed random path using TMS320 C6201, Preprint No: IS-00-21. 
Physics/0102016.\\

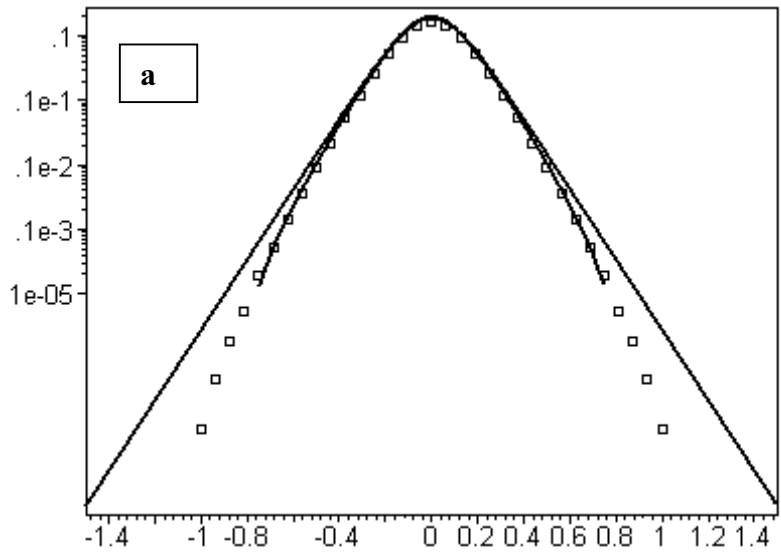
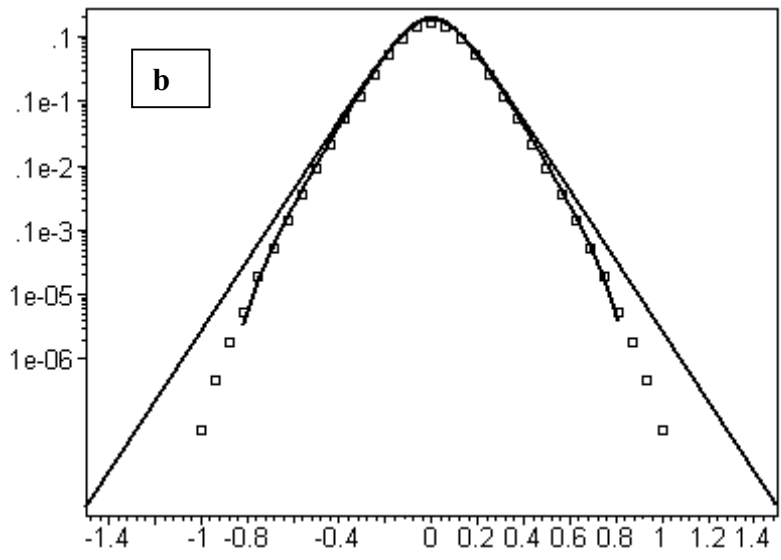
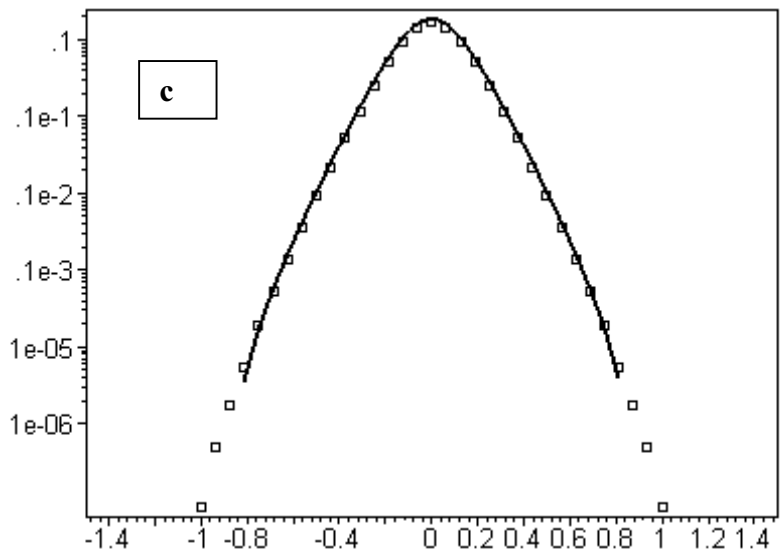
\end{document}